# Nonreciprocal second harmonic generation in a magnetoelectric material

Shingo Toyoda[1], Manfred Fiebig[1,2], Taka-hisa Arima[1,3], Yoshinori Tokura[1,4,5], and Naoki Ogawa[1,5,6]

[1] RIKEN Center for Emergent Matter Science (CEMS), Saitama 351-0198, Japan

[2] Department of Materials, ETH Zurich, Zurich 8093, Switzerland

[3] Department of Advanced Materials Science, University of Tokyo, Kashiwa 277-8561, Japan

[4]Tokyo College, University of Tokyo, Tokyo 113-8656, Japan

[5]Department of Applied Physics, University of Tokyo, Tokyo 113-8656, Japan

[6]PRESTO, Japan Science and Technology Agency (JST), 332-0012, Kawaguchi, Japan

E-mail: shingo.toyoda@riken.jp

**Abstract**

**Nonreciprocal devices that allow the light propagation in only one direction are indispensable in photonic circuits[1] and emerging quantum technologies[2]. Contemporary optical isolators and circulators, however, require large size or strong magnetic fields because of the general weakness of magnetic light-matter interactions, which hinders their integration into photonic circuits[3]. Aiming at stronger magneto-optical couplings, a promising approach is to utilize nonlinear optical processes[4,5]. Here, we demonstrate nonreciprocal magnetoelectric second harmonic generation (SHG) in $CuB_2O_4$. SHG transmission changes by almost 100% in a magnetic-field reversal of just $\pm$ 10 mT. The observed nonreciprocity results from an interference between the magnetic-dipole- and electric-dipole-type SHG. Even though the former is usually notoriously smaller than the latter, it is found that a resonantly enhanced magnetic-dipole-transition has a comparable amplitude as non-resonant electric-dipole-transition, leading to the near-perfect nonreciprocity. This mechanism could form one of the fundamental bases of nonreciprocity in multiferroics, which is transferable to a plethora of magnetoelectric systems to realize future nonreciprocal and nonlinear-optical devices.**

Unidirectional manipulation of photons is a key issue in modern information technology as exemplified by optical isolators used for lasers and in optical networks. While conventional optical isolators are the composite of a magneto-optical media, magnets, and polarizers, recent studies show that such a one-way flow of photons can be achieved in single bulk magnetoelectric materials, where time-reversal and space-inversion symmetries are simultaneously broken. For so-called magnetoelectric nonreciprocity, the nonreciprocal propagation can be characterized by a toroidal moment of the form $\vec{T} \propto \vec{P} \times \vec{M}$, where $\vec{P}$ and $\vec{M}$ are the spontaneous electric polarization and magnetization, respectively. The optical properties change with the reversal of the propagation direction of light $\vec{k}$, which can be characterized by the sign of $\vec{k} \cdot \vec{T}$ [6–8]. In the past, a wide variety of nonreciprocal phenomena has been discovered in the linear optics, including the absorption[9–12], emission[13,14], refraction[15], and diffraction[16] of light. In stark contrast, there have been few reports on the nonreciprocal phenomena in the nonlinear optical regime[17,18], although it has been naturally expected[6].

Second harmonic generation (SHG), which denotes the frequency-doubling of a light wave in a material, is one of the simplest nonlinear optical processes. SHG can be classified into two types depending on its origin. One is electric-dipole (ED) SHG, in which the frequency-doubled electric polarization $\vec{P}_i(2\omega)$ in a material is considered, while the other is magnetic-dipole (MD) SHG with the produced magnetic dipole SHG which refers to the frequency doubled magnetization $\vec{M}_i(2\omega)$ (Fig.1a). Here we neglect electric quadrupole contributions because it is forbidden for the *d-d* transition of $Cu^{2+}$ holes in $CuB_2O_4$[9]. The source term $\vec{S}(2\omega)$ of these SHG processes is given by[19]

$$\vec{S}(2\omega) = \mu_0 \frac{\partial^2 \vec{P}(2\omega)}{\partial t^2} + \mu_0 \left( \nabla \times \frac{\partial \vec{M}(2\omega)}{\partial t} \right) = -4\omega^2 \mu_0 \vec{P}(2\omega) + 4\omega\mu_0 \vec{k} \times \vec{M}(2\omega). \quad (1)$$

The SHG intensity on the detector will be $I(2\omega) \propto |\vec{S}(2\omega)|^2$, indicating that the intensity changes with the reversal of $\vec{k}$ when both processes coexist and interfere with each other (Fig.1b). Obviously, for a given $\vec{k}$, if the first term (ED-SHG) has the same amplitude and phase as the second term (MD-SHG), the total intensity $I(2\omega)$ is enhanced by the constructive interference, while the $I(2\omega)$ can be extinguished just by the reversal of $\vec{k}$. Note that we can also control the SHG interference by reversing the $\vec{P}(2\omega)$ or $\vec{M}(2\omega)$, which, for example can be realized for the latter by the reversal of the static magnetization $\vec{M}$ of the system with an external magnetic field. However, experimental realization of such a huge modulation of the SHG yield (i.e., nonreciprocal SHG of ~100% efficiency) has been a challenge. ED-SHG usually dominates MD-SHG, since in the expansion of the electromagnetic field potential with respect to $\vec{k}$, the former is of zeroth order while the latter is of first order term. This unbalance in the ED and MD transition dipole moments hampers the manifestation of large optical nonreciprocity.

Here we report on the observation of nonreciprocal SHG in $CuB_2O_4$, where the SHG intensity changes by 97 % upon the reversal of an external magnetic field. The effect is most pronounced for light around 1.4 eV, which is resonant with the intra-atomic *d-d* transition of $Cu^{2+}$ ions. We found that the MD-SHG process is resonantly enhanced at this *d-d* transition, in contrast to the ED-SHG process. The resultant resonant MD-SHG and non-resonant ED-SHG contributions are of the same order of magnitude, yielding a large nonreciprocal signal via their interference.

$CuB_2O_4$ crystalizes in a noncentrosymmetric tetragonal structure with the space-group symmetry $I\bar{4}2d$ (point group $\bar{4}2m$)[20]. The $Cu^{2+}$ ($d^9$, $S = 1/2$) ions occupy two inequivalent sites denoted as *A* and *B* sites (Fig. 1c), where $Cu^{2+}$ ions on the *A* site are responsible for the nonreciprocal optical properties[9]. This material undergoes successive magnetic transitions at $T_N = 21$ K and $T^* = 9$ K[21,22] (see Fig. 1d). Below $T^*$, magnetic moments at both *A* and *B* sites exhibit incommensurate helical order. Between $T_N$ and $T^*$, magnetic moments of the $Cu^{2+}$ ions show an easy-plane-type canted antiferromagnetic order, as schematically shown in Fig. 1e. By the application of an in-plane external magnetic field of the order of 10 mT, we can easily align the in-plane magnetization and obtain a single domain state. In the canted antiferromagnetic phase where the time-reversal and space-inversion symmetries are simultaneously broken, it shows a magnetoelectric effect which is explained by the modification of the metal-ligand hybridization with $Cu^{2+}$ moments[23,24]. The magnetically induced electric polarization appears along the *c*-axis ($P \propto \sin 2\theta$) in an external magnetic field in the *ab*-plane, where $\theta$ denotes the angle between the crystal [010] axis and the direction of the magnetic field[24]. Thus the toroidal moment $\vec{T} \propto \vec{P} \times \vec{M}$ appears in the tetragonal plane perpendicularly to the magnetic field, Fig. 1c [9,10,14,15]. One should note that the electric polarization is zero when the field is along the [100] or [010] crystal axes, thus the toroidal moment appears only when the magnetic field is tilted away from [100] or [010].

Figure 2a shows the experimental setup to detect the nonreciprocal SHG signal. The SHG intensity was measured in a transmission geometry in Voigt configuration (external magnetic field $\vec{B} \perp \vec{k}$). The light source was a regenerative amplifier, which produced 190 fs laser pulses at 6 kHz. The energy of the fundamental light was tuned by an optical parametric amplifier (OPA) to $\hbar\omega$ = 0.703 eV, for which the SHG energy $2\hbar\omega$ = 1.406 eV was resonant with the $d$-$d$ transition of $Cu^{2+}$ holes between the $d_{x2-y2}$ and $d_{xy}$ orbitals[9,22,25,26]. Note that the spectral width of the electronic transition of ~1 meV was less than the energy-spread of the incident 190-femtosecond laser pulse of ~22 meV. We used a spectrometer to resolve the fine structure of the SHG spectrum. The thickness of the sample was 50 $\mu$m with the widest crystal faces exhibiting (100) orientation, which was thin enough to match the phases between the fundamental and the SHG lights. The sample was mounted on a copper holder, which could rotate around the $c$-axis by the angle $\theta$ (Fig. 2a). Figures 2b-d show the tilt-angle dependence of the SHG contribution polarized along the $c$-axis in a magnetic-field $B$ of 50 mT. When the sample is not tilted ($\theta = 0$), the toroidal moment is zero because the electric polarization is absent for $B$ parallel to the [010] axis[24]. In this situation, the SHG signal originates solely from MD-SHG, because the complementing ED-SHG contribution with $E^{2\omega} \parallel$ [001] is zero since the associated $d$-$d$ contributions are not allowed[26]. This results in the absence of ED-MD interference and, hence, of a distinct change in SHG intensity upon the reversal of the magnetic field (see Fig. 2c). When the sample is tilted around the *c*-axis, the toroidal moment emerges along the wave vector of the light ($\vec{k} \parallel \vec{T}$). Figures 2b and 2d show the SHG spectra in the magnetic field $B = 50$ mT for the sample tilted by $\theta = -15°$ and $\theta = +15°$, respectively. The associated SHG spectra exhibit drastic changes with the reversal of magnetic field, demonstrating the reversal of the nonlinear optical nonreciprocity. Notably, the SHG spectra show a Fano-resonance-like asymmetric shape. A Fano resonance is a signature of the interference between a resonant process and a non-resonant background[27,28], in the present case represented by MD-SHG for the former and ED-SHG for the latter.

Next, we investigate the temperature dependence of the nonreciprocal SHG signal for the sample tilted by $\theta = +15°$. Figures 3a-c show the SHG spectra at different temperatures in a magnetic field $B = \pm 50$ mT. Whereas the nonreciprocal signal is clearly observed in the canted antiferromagnetic phase, it disappears in the helical and the paramagnetic phases, where the time-reversal symmetry is preserved. This result confirms that the space-and time-inversion symmetry-breaking toroidal order is essential to the emergence of the nonreciprocal behavior. The SHG signal shows broad and featureless spectra in the helical and paramagnetic phases which, referring to our analysis of the Fano-resonant behavior above, identifies its ED-SHG origin. ED-SHG can be expressed by $\vec{P}_i(2\omega) = \epsilon_0 \chi_{ijk}^{ED} \vec{E}_j(\omega) \vec{E}_k(\omega)$, where $\chi_{ijk}^{ED}$ and $\vec{E}(\omega)$ are the SHG susceptibility and the electric field of the incident light, respectively. The $\bar{4}2m$ point-group symmetry allows the SHG tensor components $\chi_{cab}$ and $\chi_{cba}$ for SHG light polarized along the *c*-axis[22,29], which becomes accessible in our experiment, where $E^\omega \perp [001]$ and $E^{2\omega} \parallel [001]$. The ED-SHG polarization $\vec{P}_c(2\omega)$ associated to $\chi_{cab}$ and $\chi_{cba}$ changes sign with the sense of rotation of the sample because the associated reversal in the sign of $\vec{E}_a(\omega)$. This explains the sign reversal in the magnetic-field dependence in Figs. 2b and 2d.

Figures 3d-f show the temperature dependence of the SHG intensity at 1.40586, 1.40656 and 1.40767 eV, which are indicated by the dashed line in Fig. 3b. At off-resonant photon energies (Figs. 3d and 3f), the ED-MD interference leads to a pronounced nonreciprocal signal in the canted antiferromagnetic phase. In contrast, the SHG intensity shows little changes with the reversal of magnetic field at the resonant energy, (see Fig. 3e).

Hence, to further elucidate the origin of the nonreciprocal signal, we investigate the SHG spectrum across a broader spectral range, (see Fig. 4a). The ED- and MD- type SHG spectra can be measured separately by properly selecting the temperature and the tilting angle $\theta$. Pure ED-SHG is obtained at 25 K in the paramagnetic phase for the sample tilted by 5 degrees, whereas pure MD-SHG spectrum is measured at 12 K in the canted antiferromagnetic phase without tilting the sample. The ED-SHG signal does not show any peak at the *d-d* transition of 1.4 eV and increases with small variations as the photon energy is raised. We ascribe this to the non-resonant virtual excitation of the charge transfer transition between $Cu^{2+}$ and $O^{2-}$ ions. On the other hand, the MD-SHG signal is resonantly enhanced at the *d-d* transitions between the $d_{x2-y2}$ and $d_{xy}$ orbitals of $Cu^{2+}$ holes at *A* site. These results explain why the ED-SHG can have an amplitude comparable to that of the MD-transition, which then leads to the strongly nonreciprocal signal.

Next, we discuss the observed spectral shape. Figure 4b illustrates a schematic of the MD- and ED-SHG susceptibilities around the resonance energy. Both the amplitude and phase of the ED-SHG contribution are almost unchanged around the resonance, because its origin is not related to the *d-d* transitions. On the other hand, phase and amplitude of the MD-SHG contribution sharply increase to show the resonance feature; the phase of the MD-SHG wave is expected to change by 180 degrees across the resonance peak. Therefore, the ED- and MD-SHG light fields interfere with constructively (destructively) each other below (above) the MD resonance, which explains the sign change of the nonreciprocal effect across the resonance energy. Notably, the amplitudes of the MD- and ED-SHG contributions can become the same at the positions indicated by dotted circles in Fig. 4b. Here we expect a perfectly nonreciprocal behavior in the SHG response by the ED-MD interference. Figure 4c shows the magnetic-field dependence of the SHG intensity at 1.40586 eV (that is, slightly below the resonance). The SHG intensity almost disappears for the negative magnetic field, whereas a strong SHG signal shows up for the same yet positive magnetic-field value, indicating that MD-SHG and ED-SHG light waves of the same amplitude yet opposite phase interfere. Slightly above the resonance (1.40767 eV), the situation is reversed, as shown in Fig. 4d. This is in excellent agreement with our model of the interference between resonant MD and non-resonant ED components.

To conclude, we experimentally demonstrate the nearly perfectly unidirectional propagation of nonreciprocal SHG light waves induced by the toroidal moment in $CuB_2O_4$. The SHG intensity changes by almost 100 % upon the reversal of a small magnetic field of 10 mT. The observed nonreciprocity originates from the interference between ED-SHG and MD-SHG. The MD-SHG process is resonantly enhanced at the *d-d* transition of $Cu^{2+}$ holes and becomes comparable in amplitude to the non-resonant broadband ED-SHG process. This work establishes that nonreciprocal wave-propagation effects are not limited to linear optics but emerge also for nonlinear optical effects and exhibit a magnitude up to the optical-diode-like unidirectional propagation of light.

## Methods

**Sample preparation**

Single crystals of $CuB_2O_4$ were grown by a flux method. Powders of CuO (7.250 g), $B_2O_3$ (15.228 g), and $LiCO_3$ (4.041 g) were mixed. They were heated in air at 1020°C, subsequently cooled down to 800°C at a rate of 1.1°C/h, and then cooled to room temperature at a rate of 16°C/h. A single crystal was oriented by using Laue X-ray diffractometer, and cut into thin plates of thickness 50 μm with the widest (100) faces. The sample surfaces specularly polished by alumina lapping films.

## Acknowledgments

S. T. is supported by Grant-in-Aid for Scientific Research from JSPS, Japan (Grant No. JP18K14154 and JP20H01867). M.F. thanks CEMS at RIKEN and ETH Zurich for supporting his research sabbatical. N.O. is supported by PRESTO JST (No. JPMJPR17I3).

## Author contribution

S.T. M.F., and T.A. designed the original project based on the idea by S.T. S.T. prepared and characterized the sample, and carried out the SHG measurements with the partial help of M.F. and N.O. All authors jointly interpreted the data and wrote the manuscript.

## Competing financial interests

The authors declare no competing financial interests.

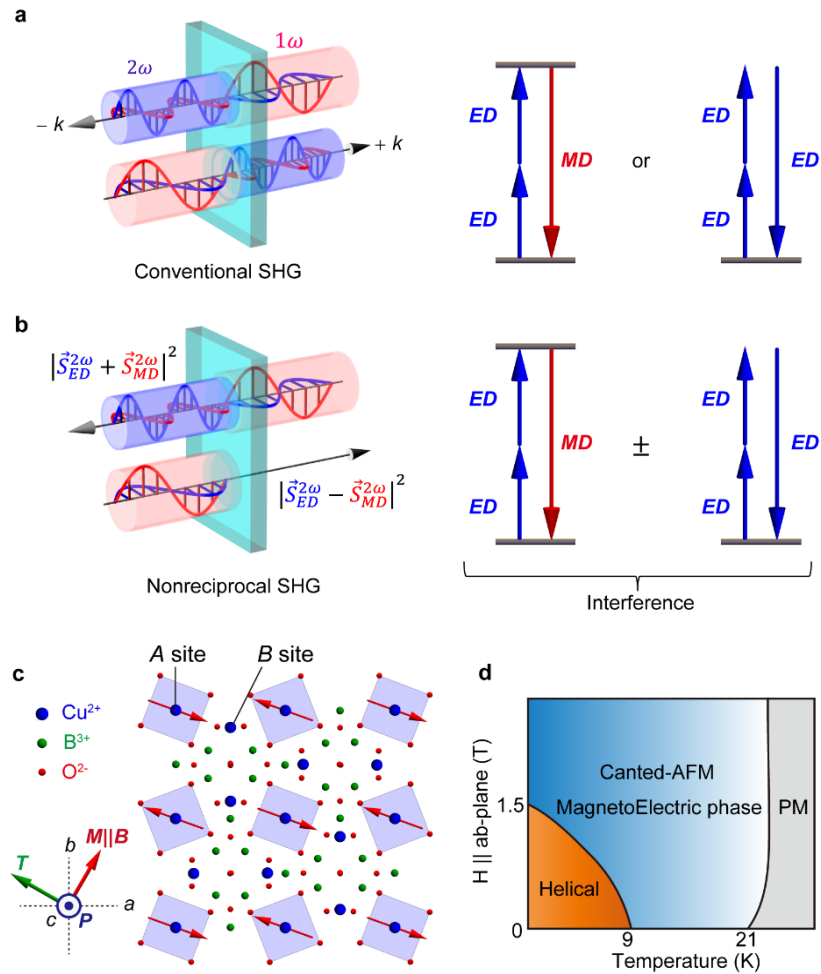

**Figure 1 | Schematic illustration of conventional SHG and Nonreciprocal SHG a,** SHG intensity generally remains unchanged after reversal of the propagation direction of the light, because the SHG signal originates from either ED-SHG or MD-SHG. Electronic transition processes of resonant MD-SHG and non-resonant ED-SHG are attached. **b,** When the ED-SHG interferes with the MD-SHG, the total intensity can depend on the propagating direction of the light. Especially, when the ED-SHG and the MD-SHG yield become similar in both amplitude and phase, their constructive interference enhances the total SHG signal for light propagating in one particular direction. The

situation reverses for the opposite propagation direction due to destructive interference, resulting in the suppression or even extinguishment of the SHG output. **c,** Crystal and magnetic structures of $CuB_2O_4$ in the canted antiferromagnetic phase projected onto the (001) plane. The red arrows indicate the magnetic moments of the $Cu^{2+}$ holes in a magnetic field $B \perp [001]$. **d,** Magnetic phase diagram of $CuB_2O_4$ in a magnetic field along the $ab$-plane. Canted-AFM and PM represent the canted antiferromagnetic and paramagnetic phases, respectively.

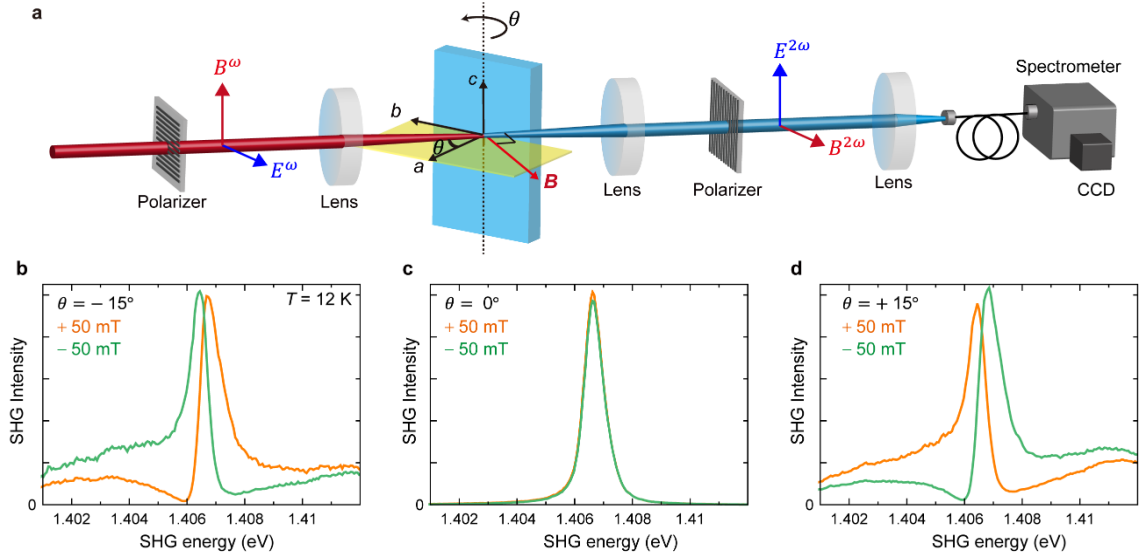

**Figure 2 | Experimental observation of nonreciprocal SHG a,** Optical setup to detect nonreciprocal SHG emission. The light pulses (photon energy $\hbar\omega$ = 0.703 eV) from an optical parametric amplifier were focused onto a CuB$_2$O$_4$ single crystal after setting their polarization to $E^\omega \perp [001]$ and $B^\omega \parallel [001]$ by using a Glan-laser prism. The sample was tilted around the *c*-axis by an angle $\theta$. SHG spectra for linearly polarized light ($E^{2\omega} \parallel [001]$ and $B^{2\omega} \perp [001]$) were measured in a transmission geometry with a spectrometer and a CCD as detector. An External magnetic field was applied in the *ab*-plane normal to the propagation direction of the light (Voigt geometry). **b-d,** Tilt angle dependence of the SHG spectra for (**b**) $\theta = -15°$, (**c**) $\theta = 0°$, and (**d**) $\theta = +15°$ measured at *T* = 12 K. Green and yellow lines show the spectra in a magnetic field of $B = -50$ mT and $B = +50$ mT, respectively.

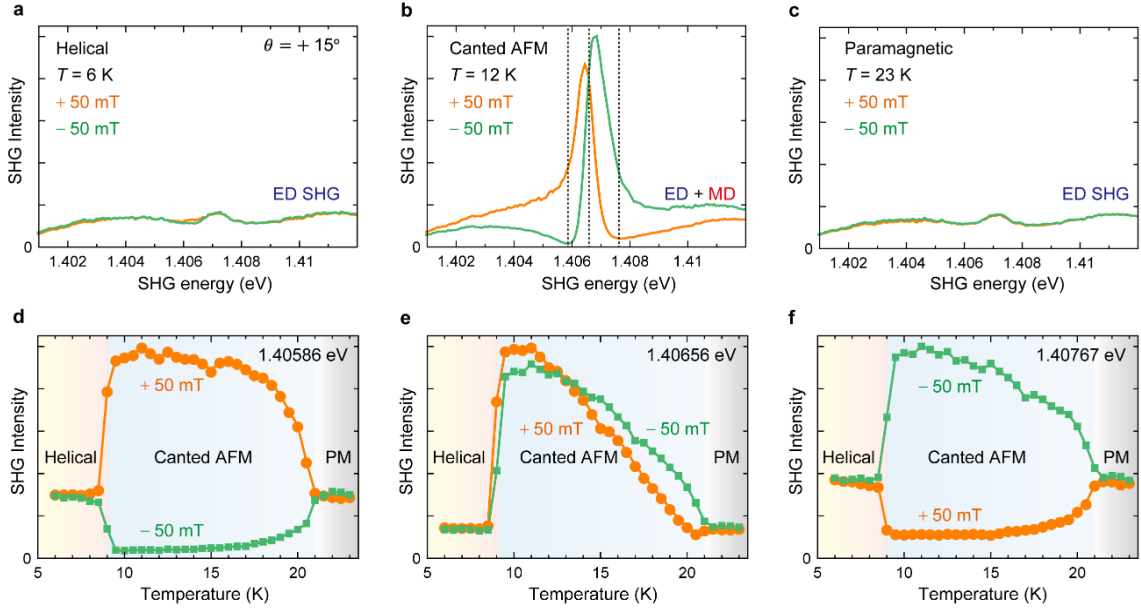

**Figure 3 | Temperature dependence of nonreciprocal SHG. a-c** SHG spectra measured at (a) 6 K in helical, (b) 12 K in the canted AFM, and (c) 23 K in the paramagnetic phase with the sample tilted by $\theta = +15°$. Green and yellow lines indicate the spectra in a magnetic field of $B = -50$ mT and $B = +50$ mT, respectively. **d-f** Temperature dependence of the SHG intensities (**d**) below the resonance energy (1.40586 eV), (**e**) at the resonance energy (1.40656 eV), and (**f**) above the resonance energy (1.40767 eV), whose photon energies are indicated by the dashed lines in **b**.

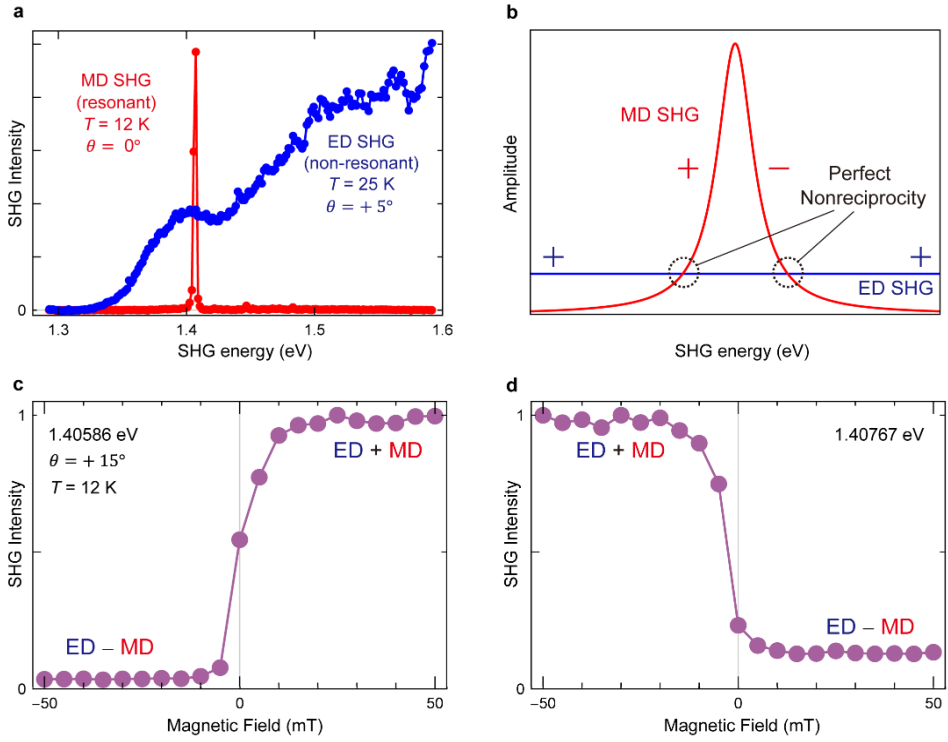

**Figure 4 | Origin of nonreciprocal SHG. a,** Spectra of the MD-SHG and ED-SHG contributions. The MD-SHG spectrum is obtained without tilting the sample ($\theta = 0°$) in a magnetic field of $B = +50$ mT at T = 12 K in the canted AFM phase. The ED-SHG spectrum is measured at $T$ = 25 K in the paramagnetic phase for a rotation angle $\theta = 5°$. Both spectra are normalized using the maximum values. **b,** Schematic of the origin of perfect nonreciprocity. The resonantly enhanced MD-SHG contribution shows a comparable amplitude as the non-resonant ED-SHG contribution, and has the same amplitude and the same (the opposite) phase below (above) the resonant photon energy as indicated by dotted circles. Blue and red signs denote the phase of the ED-SHG and MD-SHG light waves for the positive magnetic field, respectively. The former is unchanged while the

latter changes sign upon the reversal of the magnetic field. **c and d,** Magnetic-field dependence of the SHG intensity ($T$ = 12 K) for the sample tilted by +15° **(c)** below (1.40586 eV) and **(d)** above (1.40767 eV) the resonance energies. These photon energies are indicated by dotted lines in Fig. 3b.